\documentclass[12pt]{iopart}

\expandafter\let\csname equation*\endcsname=\relax 
\expandafter\let\csname endequation*\endcsname=\relax 
\usepackage{amsmath,amssymb,amsthm} 
\usepackage{braket} 

\usepackage{graphicx}
\usepackage{dcolumn}
\usepackage{bm}
\usepackage{mathtools}
\usepackage{soul}
\usepackage{color}
\usepackage{comment}
\usepackage{braket}

\makeatletter
\def\thickhline{%
  \noalign{\ifnum0=`}\fi\hrule \@height \thickarrayrulewidth \futurelet
   \reserved@a\@xthickhline}
\def\@xthickhline{\ifx\reserved@a\thickhline
               \vskip\doublerulesep
               \vskip-\thickarrayrulewidth
             \fi
      \ifnum0=`{\fi}}
\makeatother

\newlength{\thickarrayrulewidth}
\begin{document}

\title{ Mass-coupled relativistic spontaneous collapse models}

\author{C. Jones}
 \address{Department of Physics, University of Trieste,
Strada Costiera 11, 34151 Trieste, Italy
}%
 \address{Istituto Nazionale di Fisica Nucleare, Trieste Section,  Via Valerio 2, 34127 Trieste, Italy}
  \ead{caitlinisobel.jones@phd.units.it}
  
\author{G. Gasbarri}%
\address{F\'isica Te\`orica: Informaci\'o i Fen\`omens Qu\`antics, Department de F\'isica, Universitat Aut\`onoma de Barcelona, 08193 Bellaterra (Barcelona), Spain}
\ead{g.gasbarri@uab.cat}

\author{A. Bassi}
 \address{Department of Physics, University of Trieste,
Strada Costiera 11, 34151 Trieste, Italy\
}%
 \address{Istituto Nazionale di Fisica Nucleare, Trieste Section,  Via Valerio 2, 34127 Trieste, Italy}

\date{\today}            
      
\begin{abstract}	
	Currently there is not a satisfactory relativistic spontaneous collapse model. 
Here we show the impossibility of a simple generalization of the continuous spontaneous collapse (CSL) model to the relativistic framework. We consider a mass coupled model as in the non-relativistic limit this gives the CSL model.
	We show that a Lorentz covariant collapse equation cannot simultaneously satisfy the following conditions:
	i) To avoid a divergent rate of energy,
	ii) To prevent superluminal signaling.
	
\end{abstract}

\maketitle   

\section{Introduction}
The transition between the quantum and classical realm has long been a problem faced by quantum mechanics \cite{bell2001there}. 
Quantum theory describes microscopic objects which can be found in superpositions and classical mechanics describes macroscopic objects which cannot. Explaining the transition between these two regimes has been a problem quantum mechanics has faced since its inception, sometimes called the macroscopic objectification problem \cite{busch1991problem}.

Spontaneous collapse models \cite{ghirardi1986unified,pearle1976reduction, csl, bassi2003dynamical,diosi1987universal,diosi1989models,bassi2013models} are an attempt to solve this problem by modifying the Schr\"odinger equation to give a non-unitary and stochastic time evolution. 
 This modification has two important properties. The first is that it causes superpositions of the state to spontaneously collapse in a particular basis. Since macroscopic objects are never found in spatial superpositions, most collapse models  localise the state in the position basis. The second property is that the rate of collapse scales with the mass, and this is known as the amplification mechanism.
It is exactly this mechanism that guarantees that collapse models recover the quantum mechanical behaviour for microscopic systems, and the classical behaviour for macroscopic objects.    

Since their introduction  there have been a plethora of different collapse models proposed, see \cite{bassi2003dynamical} and references within. As mentioned above most of these models cause collapse in the position basis and hence are coupled to the mass or number density operator. It is this class of spontaneous collapse models will we study in this paper. 
   Continuous Spontaneous Localisation (CSL) \cite{csl}  and its non-white noise generalisation coloured CSL \cite{adler2007collapse} are the most studied of such mass-coupled models. In this class, another spontaneous collapse model of note is the Di\'osi-Penrose (DP)  model \cite{diosi1987universal,penrose1996gravity} which proposes gravity induced wavefunction collapse. As they offer predictions which differ from conventional quantum mechanics, there are experimental projects aimed at testing the validity of these theories \cite{piscicchia2017csl,carlesso2016experimental,vinante2017improved}.  Much of the possible parameter space for these models has been explored, and research is still ongoing. 
    
    As originally formulated neither CSL, coloured CSL nor the DP model are compatible with special relativity. In order to be fundamental, a spontaneous collapse model is expected to be in agreement with special relativity, assuming special relativity is correct.  There have been many attempts to construct a relativistic collapse model \cite{breuer1998relativistic,breuer1999stochastic,pearle1999relativistic,nicrosini2003relativistic,tumulka2006relativistic,bedingham2011relativistic,bedingham2014matter,tilloy2017interacting,bedingham2019csl,tumulka2020relativistic}, but none of them proved to be satisfactory. 
Every model so far proposed has encountered some issues, for example having a divergence in rate of change of energy \cite{pearle1999relativistic} or not being able to describe indistinguishable particles \cite{breuer1998relativistic,tumulka2006relativistic,tumulka2020relativistic} etc.  

The purpose of this work is to understand why these attempts at relativistic collapse models fail. This is done by starting from the most general form of a continuous time spontaneous collapse model and then applying a set of minimal requirements for a successful relativistic collapse model. We will  consider the case of scalar bosons but the results of this work are not expected to differ for fermions. We assume that state vector reduction occurs in the mass density basis. These requirements are; that the map is Lorentz covariant, that the rate of change of energy is finite, and that microcausality is respected. It is found that it is not possible for all the requirements to be simultaneously satisfied.  We also find that the requirement for finite energy ensures that the theory must have a well defined non-relativistic limit.

This work is laid out as follows: in section \ref{general_form} the general structure of a spontaneous collapse model is given and it is shown how the dynamics is characterised by a collapse operator and a two point correlation function. In section \ref{lorentz_covariance} it will be shown that the requirements for this map to be Lorentz covariant are that given the collapse operator is a Lorentz scalar, the two point correlation function is a function of the invariant 4 distance.
The collapse operator is specified in  section \ref{mass_coupled}.  In section \ref{finite_energy} it is shown that the rate of change of the energy under the map is only finite if  the two point correlation function is zero in the limit of the two points coinciding. Finally in section \ref{sls_micro} it is shown how under the above requirements such a dynamics permits superluminal signalling.

\section{General Form of a Spontaneous Collapse Model} \label{general_form}
Spontaneous collapse models are described by non-linear stochastic Schr\"odinger equations with the usual deterministic evolution associated with the unitary dynamics and a non-linear term coupled to a gaussian stochastic noise field~\cite{adler2007collapse,bassi2003dynamical,ghirardi1986unified}.
For the purposes of this article however it is only necessary to consider the evolution of the statistical operator $\hat{\rho}$ which is averaged over the stochastic noise.  Throughout this work we set $\hbar = c= 1$.
For a collapse models the evolution of $\hat{\rho}$ is given by\cite{gasbarri2018stochastic,gasbarri2017gravity,adler2007collapse}:
\begin{align}
\hat{\rho}_{t}= \mathcal{U}_{t}^{0}[\mathcal{M}_{t}[\hat{\rho}_{0}]]
\label{nm_gen}
\end{align}
where $\mathcal{U}_{t}^{0}[\cdot] = e^{-i\hat{H}t}\cdot e^{i\hat{H}t} $ is the standard quantum evolution, with $\hat{H}$ as the standard free Hamiltonian and $ \mathcal{M}_t $ is the contribution due to the presence of the stochastic noise and has the following expression
 \begin{equation}
 \label{map_form_general}
 \mathcal{M}_t  =\overleftarrow{\text{T}} \exp \left( \gamma {\mathcal{L}_t}\right) 
 \end{equation}
 with
 \begin{equation}
 \label{nm_map_general} 
\mathcal{L}_{t} \!= \!\!\int_{\Omega_{t}}\!\!\! d^4 x\! \!\int_{\Omega_{t}}\!\!\!d^4 y  D(x,y)
\big[ \hat{Q}^{L}(y)\hat{Q}^{R}(x) - \theta(x^0 - y^0)\hat{Q}^{L}(x)\hat{Q}^{L}(y)
- \theta(y^0 - x^0) \hat{Q}^{R}(y)\hat{Q}^{R}(x) \big]
\end{equation}
 $\hat{Q}(\mathbf{x})$ is an arbitrary self-adjoint operator, $\hat{Q}(x)= \mathcal{U}_{x^0}^{0 \dagger}[\hat{Q}(\mathbf{x})]$,  $\Omega_{t}= \{ x = (\mathbf{x},x^0)  \, |\, x^0 \in [0,t] ,\,  \mathbf{x} \in \mathbb{R}^{3}\} $, the superscript L (R) denotes operator acting on the statistical operator $\hat{\rho}$ from left (right), i.e. $\hat{Q}^R(x) \hat{Q}^L(x)  \hat{\rho} =  \hat{Q}(y)  \hat{\rho} \hat{Q}(y)$ and $\overleftarrow{\text{T}}$ is the chronological time ordering acting on the L/R operators and is defined as 
 \begin{eqnarray}Q^{L/R}(y)Q^{L/R}(x) = 
 \left\{ \begin{array}{c}
 Q^{L/R}(x)Q^{L/R}(y) \,\, \text{if}\,\, x^0 > y^0\\
Q^{L/R}(y) Q^{L/R}(x) \,\, \text{if}\,\, y^0 > x^0
 \end{array}\right.
 \end{eqnarray} 
 Equation  \ref{map_form_general} suppresses the  off diagonal elements of $\hat{\rho}$ in the basis of the eigenstates of $\hat{Q}(\mathbf{x})$.
 For collapse models this is a consequence of the collapse of the wavefunction, which again occurs in the eigenbasis of $\hat{Q}(\mathbf{x})$ when the free Hamiltonian dynamics is neglected, see \cite{gasbarri2017gravity} for a proof.
 It is for this reason that many collapse models have  $\hat{Q}(\mathbf{x})$ as the mass density or number density operator, since this means macroscopic objects do not remain in spatial superposition.

As one can see from Eq.~\eqref{nm_map_general}, all collapse models are defined by specifying $D(x,y)$,  $\hat{Q}(x)$, and the coupling $\gamma$.
  For example the average map of the coloured CSL model is characterized by the above equation with:
	\begin{align}
		\label{c_csl_master}
	D(x,y)
&=   \frac{1}{m_0^2(4 \pi r_C^2)^{3/2}} \exp[-(\mathbf{x}- \mathbf{y})^2/ 2r_C^2] F(t-s) \nonumber \\
	\hat{Q}(x) & = \hat{M}(x)
	\end{align}
where $F(t-s)$ is an arbitrary positive function of time, $\hat{M}(x)$ is the non-relativistic mass density operator, $m_{0}$ a reference mass (usually the nucleon mass),  and $r_C$ is a parameter of the model which determines the length scale of the collapse. Original CSL is the recovered when $F(t-s) = \delta(t-s)$.  This choice of $\hat{Q}(x)$ ensures that the wavefunction is driven to collapse in the mass density basis, which correctly predicts that macroscopic objects are not found in spatial superpositions. The form of $D(x,y)$ means that the wave-function is localised by a Gaussian of width $r_C$ where the amplitude of the wavefunction is maximised, which means that the theory effectively replicates the Born probability rule. The standard CSL model is Galilei invariant \cite{bassi2003dynamical}, guaranteeing that, in a non-relativistic setting, the dynamical law are the same in all inertial frames.

\section{Lorentz Covariance of Non-unitary Dynamics}  
\label{lorentz_covariance}
In a probabilistic theory consistent with special relativity the transition probabilities  between point-like events  in spacetime is Poincar\'e covariant.
For this work it is sufficient to check whether the conditional probability distribution is invariant under the Lorentz group, instead of the larger Poincar\'e group. So from on we will consider only Lorentz group transforms. We can reasonably do this as this work shows that a Lorentz covariant CSL model is not possible, and as Poincar\'e covariance is a strictly stronger condition it follows that a Poincar\'e covariant CSL model is also not possible. Furthermore we will consider only the orthochronous Lorentz group as the dynamics is for positive time translations only.

So for a theory to be consistent with special relativity the transition probability between point-like events in spacetime

\begin{equation}
\label{general_cond_prob}
P(A(x_1), B(x_2)\, ...| C(y_1), D(y_2)\, ...)
\end{equation}
 must be invariant under Lorentz transformations, where $A(x_1)$ is an event at position $x_1$ in space-time and so on.
For the case of collapse models the covariance of the transition probabilities is guaranteed only if the dynamics of the statistical operator is covariant, as already noticed in~\cite{bassi2003dynamical, miller2012sixty}.
Since the map $\mathcal{U}_{t}^{0}$ describing the standard free quantum evolution is Lorence convariant we are left to verify the covariance of the collaspe contribution $\mathcal{M}_{t}$.   
Therefore, in the rest of the paper we will work with the map $\mathcal{M}_{t}$.

We call the map $\mathcal{M}_{t}$ Lorentz covariant if:
\begin{equation}
\label{lorenz_cov}
 \mathcal{U}_{\Lambda}  \mathcal{M}_{t} \mathcal{U}_{\Lambda}^{-1} = \mathcal{M}_{\Lambda(t)},
\end{equation}
where  $ \mathcal{U}_{\Lambda}[\cdot]= \hat{U}(\Lambda)\,\cdot\, \hat{U}(\Lambda)^{\dagger}$  and $\hat{U}(\Lambda)$ is the standard unitary representation of Lorentz group on the Hilbert space of the system and $\Lambda$ is an element of the orthochronous Lorentz group \cite{davies1970repeated}.
Then with the help of Eq.~\eqref{map_form_general} and  Eq.~\eqref{nm_map_general} and by exploiting the unitarity of $ \mathcal{U}_{\Lambda}$, it is straightforward to show that  Eq.~\eqref{lorenz_cov}  is satisfied if and only if the following condition is satisfied:
 \begin{align}
 \label{lorenz_cov_expli_2}
 &\overleftarrow{\text{T}}\exp\Bigg\{\int_{\Omega_{t}} d^4 x \int_{\Omega_{t}} d^4 y  D(x,y) 
\Big[\mathcal{U}_{\Lambda}[\hat{Q}^L(y)] \mathcal{U}_{\Lambda}[\hat{Q}^R( x)]\nonumber\\
&\hspace{1,4cm} - \theta (x^0 - y^0)  \mathcal{U}_{\Lambda}[\hat{Q}^L(x)]\mathcal{U}_{\Lambda}[\hat{Q}^L( y )] 
 - \theta(y^0 - x^0)  \mathcal{U}_{\Lambda}[\hat{Q}^R(y)]\mathcal{U}_{\Lambda}[\hat{Q}^R(x)]\Big]\Bigg\}\nonumber\\
 &=\overleftarrow{\text{T}}\exp\Bigg\{\int_{\Lambda(\Omega_{t})}\!\!\!\!\!d^4 x \int_{\Lambda(\Omega_{t})}\!\!\!\!\! d^4 y  D(x,y)  \nonumber\\
&\hspace{1,4cm}  \Big[\hat{Q}^L(y )\hat{Q}^R(x)- \theta (x^0 - y^0)  \hat{Q}^L(x)\hat{Q}^L( y)- \theta(y^0 - x^0)  \hat{Q}^R(y)\hat{Q}^R(x)\Big]\Bigg\} 
 \end{align}

Let us consider the case when the operator $\hat{Q}$ is a Lorentz scalar:
  \begin{equation}
 \label{lorentz_scalar}
  \mathcal{U}_{\Lambda}[\hat{Q}(y)]  = \hat{Q}(\Lambda(y)).
 \end{equation}
In this case condition in Eq.~\eqref{lorenz_cov_expli_2} is met only if  $D(x,y) = D(|x-y|)$~\cite{ticciati1999quantum}, in which case $D(x',y')=D(x,y)$,
 where $x' = \Lambda (x)$.

Finally we note that as $ D(x,y)$  is a function of the 4-distance, then it is symmetric under exchange of $x$ and $y$. This allows Eq.~\eqref{nm_map_general} to be simplified to:
\begin{equation}
\label{rearranged_generator}
\mathcal{L}_t  = \int_{\Omega_{t} }  \!\!\! d^4 x \int_{\Omega_{x^0}} \!\!\! d^4y \, D(x,y)
\big[\hat{Q}^L(y) \hat{Q}^R(x) + \hat{Q}^L(x)\hat{Q}^R(y) -\hat{Q}^L(x) \hat{Q}^L(y) -\hat{Q}^R(x)\hat{Q}^R(y)\big] 
\end{equation}
which is a form which makes calculations easier to perform.

\section{Choice of the collapse operator} \label{mass_coupled}
As discussed in Section II, the choice of the operator $\hat{Q}(x)$ determines the basis that spontaneous collapse  occurs in. 
We seek a model that is relativistically invariant and at the same time is able to reproduce the CSL dynamics in the non relativistic limit. A natural choice is to select $\hat{Q}(x)$ to be the relativistic mass density operator, i.e.
\begin{equation}
\label{rela_number density} 
\hat{Q}(x) = m \hat{a}^\dagger(x) \hat{a}(x)
\end{equation}
where $\hat{a}^\dagger(x)$ is the the positive energy part of the  Klein-Gorden field operator. From now on we take Eq.~\eqref{rela_number density} to be the collapse operator.

Previous attempts to derive relativistic extensions of CSL model have encountered the problem of an infinite energy rate in the dynamics.
In the next section we will find a necessary condition to avoid this unpleasant feature from a relativistic collapse model.

	\section{Finiteness of rate of change of energy} \label{finite_energy}
	An important physical requirement for the dynamics to satisfy is that the rate of change of energy is finite. Some previous attempts to find a relativistic collapse model have suffered from  this divergence \cite{bedingham2019csl,bassi2003dynamical}. 	
	Here we show that the dynamics given by Eq.~\eqref{nm_map_general} with the choice of  Eq.~\eqref{rela_number density}  as the collapse operator does not lead to a divergent rate of change of energy provided that the correlator prevents large transfers of momentum between momentum eigenstates. 
	 We show this for a single particle system in the weak coupling regime.
	 
	The rate of change of the energy  is given by:
	\begin{equation}
	\label{roc_approx}
	\frac{d}{dt}  \text{Tr}  ( \hat{H} \mathcal{M}_t [\hat{\rho}]) = \frac{d}{dt} \text{Tr}  (\hat{H} \: \overleftarrow{\text{T}} \exp {\Lambda \mathcal{L}_t [\hat{\rho}]} ),
	\end{equation}
	where $\hat{H}$ is the Hamiltonian for the free dynamics \cite{peskin2018introduction}:
	\begin{equation}
	    \label{H_0} 
	    \hat{H} = \int \frac{d\mathbf{p}}{(2\pi)^{3}} \:  E_{\mathbf{p}} \hat{a}^\dagger_{\mathbf{p}} \hat{a}_{\mathbf{p}} 
	\end{equation}
	where $E_{\mathbf{p}} =\sqrt{m^2 + \mathbf{p}^2}$ and $\mathbf{p}$ is the three momentum and  $\hat{a}_{\textbf{p}} $  and $\hat{a}^\dagger_{\textbf{p}}$ are the creation and annihilation operators in momentum eigenbasis and satisfy the following commutation relation:
	 \begin{equation}
	 \label{commute_defn}
	 [\hat{a}_{\textbf{p}} , \hat{a}^\dagger_{\textbf{q}}] =   (2\pi)^{3}\delta^3(\textbf{p} -\textbf{q})
	 \end{equation}
	 We expand the map up to the first order in $\gamma$, to obtain:
	\begin{equation}
	\label{order_expand}
	\overleftarrow{\text{T}} \exp \gamma \mathcal{L}_t [\hat{\rho}] =  \mathbb{I} + \gamma  \mathcal{L}_t [\hat{\rho}] + \mathcal{O}(\gamma^2).
	\end{equation}
	Exploiting this and dropping terms of orders higher than $\gamma$ gives:
	\begin{equation}
	\label{roc_energy}
	\frac{d}{dt}  \text{Tr}  ( \hat{H} \mathcal{M}_t [\hat{\rho}]) \approx \gamma  \text{Tr}   ( \hat{H}  \mathcal{L}_t [\hat{\rho}] )
	\end{equation}
We evaluate Eq.  \eqref{roc_energy} in the single particle sector, where the state of the system can be expressed as:

	 \begin{equation}
	 \label{single_particle_rho} 
	 \hat{\rho} = \int d\mathbf{p} \int d\mathbf{q} \: A(\mathbf{p},\mathbf{q}) \: \hat{a}^\dagger_{\textbf{p}} |0\rangle \langle 0|\hat{a}_{\textbf{q}}
	 \end{equation}
 where $A(\mathbf{p}, \mathbf{q})$ is an arbitrary function such that:
 \begin{equation}
 \label{normalised_cond}
 \text{Tr} (\hat{\rho}) =  \int d\mathbf{p} \: A(\mathbf{p},\mathbf{p})  = 1
 \end{equation}

	In order to calculate  Eq.~\eqref{roc_energy} we expand the correlation function and the collapse operator in Fourier components, i.e.
	\begin{align}
	\label{ft_funcs}
	&D(\mathbf{x},t) =  \frac{1}{(2 \pi)^3}\int d\mathbf{q}\, e^{i\mathbf{x}\cdot\mathbf{q}}\tilde{D}(\mathbf{q},t)  \nonumber \\
	&\hat{Q}(\mathbf{x},t)  = \frac{m}{(2 \pi)^6} \int d\mathbf{q} \int d\mathbf{p} \; \frac{1}{2\sqrt{ E_{\mathbf{p}} E_{\mathbf{q}} }} \, e^{i\mathbf{x}\cdot (\mathbf{q} -\mathbf{p}  )} e^{-i( E_{\mathbf{q}} - E_{\mathbf{p} }) t } \hat{a}^{\dagger}_{\mathbf{q}}\hat{a}_{\mathbf{p}}
	\end{align}
	substituting these into Eq.~\eqref{rearranged_generator} with $D(x,y) = D(|x-y|)$ and then integrating over the spatial variables $d\mathbf{x}$ and $d\mathbf{y}$ lets  $\mathcal{L}_{t}$ be written as:
{
	\begin{align} \label{k_l}
\mathcal{L}_{t}=\frac{1}{(2\pi)^{12}}\int \!\! d\mathbf{q} \int_{0}^{t} d s \int_{0}^{s} d\tau\, \tilde{D}(\mathbf{q},s-\tau )&\big\{
\hat{K}^{L}(\mathbf{q},s)\hat{K}^{\dagger,L}(\mathbf{q},\tau)+\hat{K}^{R}(\mathbf{q},s)\hat{K}^{\dagger,R}(\mathbf{q},\tau) \nonumber \\&
-\hat{K}^{\dagger,L}(\mathbf{q},\tau)\hat{K}^{R}(\mathbf{q},s)-\hat{K}^{L}(\mathbf{q},s)\hat{K}^{\dagger,R}(\mathbf{q},\tau)\big\}
\end{align}
}
	where:
	\begin{align}
	\label{K_operator}
	\hat{K}(\mathbf{q},t) = m \int \frac{d\mathbf{p}}{2\sqrt{ E_{\mathbf{p}} E_{\mathbf{p}-\mathbf{q}}}}  e^{i\Delta E(\mathbf{p},\mathbf{q})t} \hat{a}^{\dagger}_{\mathbf{p}}\hat{a}_{\mathbf{p}-\mathbf{q}}
	\end{align}
	with $\Delta E(\mathbf{p},\mathbf{q})= E_{\mathbf{p-q}}-E_{\mathbf{p}}$.
	Note that $\hat{K}(\mathbf{q},t)=\hat{K}^{\dagger}(-\mathbf{q},-t)$.
	
	Then using Eq.~\eqref{k_l} and Eq.~\eqref{ft_funcs} we have that Eq.~\eqref{roc_energy} evaluates to: 
	    \begin{align}
	    \label{finite_energy_final}
 &\text{Tr}  ( \hat{H}  \mathcal{L}_t [\hat{\rho}] )  =   \nonumber\\
 &-\frac{m^{2}}{(2\pi)^{3}}\int\! d\mathbf{q}\! \int\! d \mathbf{p}\! \int_{0}^{t}\!\!ds\! \int_{0}^{s}\!\! d\tau  \tilde{D}(\mathbf{q},s-\tau) \cos(\Delta E(\mathbf{p},\mathbf{q})( s-\tau))A(\mathbf{p},\mathbf{p})\left( \frac{1}{2E_{\mathbf{p}-\mathbf{q}}}-\frac{1}{2E_{\mathbf{p}}}\right) \qquad
	    \end{align} 
	We are interested to see if this integral is finite.  
	The integrals over both $\textbf{q}$ and $\textbf{p}$ are over an infinite range,
	but if the system is assumed to initially have a finite energy then the integral over $\mathbf{p}$ will automatically converge to a finite value. Then in order for the integral over $\mathbf{q}$ to converge a necessary condition is that:
	\begin{equation}
	\label{cutoff_condition} 
   \lim_{|\mathbf{q}| \to \infty} \tilde{D}(|\mathbf{q}|,s-\tau) = 0.
	\end{equation}
	
	 Note here that if the noise is white, i.e. if $D(x,y) = \delta^4 (x-y)$ then:
	 \begin{equation}
	    \tilde{D}(|\mathbf{q}|,s-\tau) = (2 \pi)^3\delta(s-\tau)
	 \end{equation}
and the energy rate in Eq.~\eqref{finite_energy_final} diverges.  Therefore, relativistic white-noise models are physically inconsistent, as already noticed in 
chapter 13 of \cite{bassi2003dynamical} and \cite{bedingham2019csl}.
 Through a similar calculation it can be seen that if the correlation function $D(x-y)$ further satisfies the  following condition 
\begin{equation}
\label{cutoff_condition_non_real}
\tilde{D}(|\mathbf{q}|,t) \approx 0 \quad \text{for}  \: \:\mathbf{q} > \kappa m
\end{equation}
where $\kappa \gg 1$ is a fixed constant, then the map described by Eq.~\eqref{nm_map_general} is well behaved in the non relativistic sector, i.e. it leaves non-relativistic particles non relativistic. 
	 This is shown in~\ref{non-rela}.
Notice that  condition Eq.~\eqref{cutoff_condition} is a weaker condition than Eq.\eqref{cutoff_condition_non_real}, hence if the model has a NR limit then it also has a finite energy rate.

\section{Superluminal Signalling and Micro-causality} \label{sls_micro} 

  We have given the minimal requirements for the map in Eq.~\eqref{rearranged_generator} to guarantee that the wave-function collapses and  has a finite energy rate. Here we check conditions under which the model does not allow superluminal signalling by checking if the model satisfies the microcausality condition.
 We will show that given the requirements from sections \ref{lorentz_covariance}, \ref{mass_coupled} and \ref{finite_energy} the dynamics described by Eq.~\eqref{rearranged_generator}  violates  microcausality. For this section we will work in the Heisenberg picture, therefore operators evolve with the dual map $\mathcal{M}_{t}^{*}[\mathcal{U}^{0*}_{t}[\,\cdot\,]]$.

In standard quantum mechanics where the dynamics $\mathcal{U}_{t}$ is unitary the microcausality condition reads
\begin{equation}
 \label{mirco_causal_general}
 [\hat{A}(z_1), \hat{B}(z_2)] = 0 \hspace{2cm}\forall \hspace{0.3cm} |z_1-z_2|< 0
 \end{equation}
where $\hat{A}(z_1)=\mathcal{U}_{t_1}^{0*}\hat{A}[(\mathbf{z}_1, 0)]$ and $\hat{B}(z_2)= \mathcal{U}_{t_2}^{0*}[\hat{B}(\mathbf{z}_2,0)]$ are local operators, and $z_1$ and $z_2$ are two points in spacetime such that  $z_1 =(\mathbf{z}_1, t_1)$ and  $z_2 =(\mathbf{z}_2, t_2)$ with $\mathbf{z}_1$ and $\mathbf{z}_2$ are two points in 3D space. The above condition is synonymous with no superluminal signalling, because it guarantees that local measurements at space-like separated points do not influence each other.
 For the case of a non unitary dynamics it is not possible to evaluate  Eq.~\eqref{mirco_causal_general} 
 because:  
 \begin{align}
 \label{nm_failed_evo_equal_time}
 & \mathcal{M}^*_{t_1}[\mathcal{U}_{t_1}^{0*}\hat{A}(\mathbf{z}_1, 0) \mathcal{U}_{t_1}^{0*}\hat{B}(\mathbf{z}_2, 0)]  \neq \mathcal{M}^*_{t_1}[\mathcal{U}_{t_1}^{0*}\hat{A}(\mathbf{z}_1, 0)] \mathcal{M}^*_{t_1}[\mathcal{U}_{t_1}^{0*}\hat{B}(\mathbf{z}_2, 0)]. 
 \end{align} In words, for non-unitary dynamics the evolution of the product of two operators $\hat{A}$ and $\hat{B}$ cannot be described by the product of the independently evolved operators.

One way around this problem is if the non unitary map admits a unitary stochastic unravelling $\tilde{\mathcal{U}}_{t}$, i.e. $\mathbb{E}(\tilde{\mathcal{U}}_t) = \mathcal{U}_{t}^{0}\mathcal{M}_{t}$  as in the case of the map in Eq.~\eqref{rearranged_generator}
(for an explicit expression of the unitary stochastic unravelling see~\cite{adler2007collapse,bassi2003dynamical}).
\\In this case one could define a microcausality condition as:
\begin{align}
\label{mc_cond_unit_unravelling}
&\mathbb{E}([\hat{A}(z_1), \hat{B}(z_2)])  =0 \hspace{2cm}\forall \hspace{0.3cm} |z_1-z_2|< 0
\end{align} 
where $ \hat{A}(z_1) \equiv \mathcal{\tilde{U}}^{*}_{t_1}[\hat{A}(\mathbf{z}_1, 0)]$ and $ \hat{B}(z_2)\equiv \mathcal{\tilde{U}}^{*}_{t_2}[\hat{B}(\mathbf{z}_2,0)]$ are local operators. 

However the presence of the stochastic average makes the equation hard to evaluate in full generality, but for our purpose it is sufficient to restrict the study to the less general case $\hat{A}=\hat{B}=\hat{\phi}(x)$, $t_1=t$ and $t_{2}=0$. Under these assumptions  
  Eq.~\eqref{mc_cond_unit_unravelling} reduces to:
  \begin{align}
  \label{special_case}
  & \mathbb{E}([ \tilde{\mathcal{U}}_{t}^{*}[ \hat{\phi}(\mathbf{z}_1, 0)], \hat{\phi}(\mathbf{z}_2, 0)])
   =  [\mathcal{M}^*_{t}[\hat{\phi}(\mathbf{z}_1, t)],  \hat{\phi}(\mathbf{z}_2, 0)] =0 
  \end{align}
  where  $\hat{\phi}(\mathbf{z}_1, t)= \mathcal{U}_{t}^{0*}[\hat{\phi}(\mathbf{z}_1, 0)]$ is the field operator evolved under the standard free evolution dynamics. 
We expand the map $\mathcal{M}^*_{t}$ to first order in $\gamma$. By substituting the Fourier expansion for $\hat{\phi}(x)$ (see~\ref{mc_calc}) we arrive at:
 \begin{align}
	\label{perturbed_no_sl_calc}
&[\mathcal{M}_t^* [\hat{\phi}(\mathbf{z}_{1}, t)], \hat{\phi}(\mathbf{z}_{2},0)] \approx   [  \hat{\phi}(\mathbf{z}_{1},t) + \gamma \mathcal{L}_t^* [\hat{\phi}(\mathbf{z}_{1},t)], \hat{\phi}(\mathbf{z}_{2},0)] \nonumber   \\ 
 &\hspace{0.5cm}= [\hat{\phi}(\mathbf{z}_{1},t), \hat{\phi}(\mathbf{z}_{2},0)] 
 +\gamma \int_{\Omega_{t}} \!\!\! d^4 x   \int_{\Omega_{x^0}} \!\!\!\! d^4 y \, D(x-y ) [[\hat{Q}(y), [\hat{\phi}(\mathbf{z}_1,t), \hat{Q}(x)]],\hat{\phi} (\mathbf{z}_2,0)]\nonumber\\
&\hspace{0.5cm}=[\hat{\phi}(\mathbf{z}_{1},t), \hat{\phi}(\mathbf{z}_{2},0)] 
 +\gamma m^{2} \!\!\int_{0}^{t}\!\! d\tau \!\! \int d\mathbf{x}
 D(\mathbf{x}, \tau)\left\{F(\mathbf{z}_{1}-\mathbf{z}_{2},\mathbf{x}, t, \tau)-F(\mathbf{z}_{2}-\mathbf{z}_{1},-\mathbf{x},t, \tau)\right\}
\end{align}
where
\begin{align}
\label{f_def}
F(\mathbf{z}_{1}-\mathbf{z}_{2},\mathbf{x}, t, \tau)=
    \int\frac{d\mathbf{k}}{2E_{\mathbf{k}}}\!\int\!\frac{d\mathbf{k'}}{4E_{\mathbf{k'}}^{2}} e^{i[\mathbf{k}\cdot(\mathbf{z}_{1}-\mathbf{z}_{2})-E_{\mathbf{k}}t]} e^{-i[(\mathbf{k}-\mathbf{k}')\cdot \mathbf{x}-(E_{\mathbf{k}}-E_{\mathbf{k'}})\tau]}
\end{align}
 The first term is the  microcausality condition for standard quantum fields while the second term is due to the non-unitary evolution.
From this equation one immediately notices that the microcausality condition could be satisfied only if:
1. $D(x)=0$, i.e.  the dynamics is unitary and no collapse mechanism is present, but this is not the working hypothesis of this paper; 
2. $F(\mathbf{z}_{1}-\mathbf{z}_{2},\mathbf{x},t, \tau) =  F(\textbf{z}_2 -\textbf{z}_1, -\mathbf{x},t, \tau)$, but a simple inspection at Eq.~\eqref{f_def} shows that in general this is not true
3. $D(x-y)=\delta^{(4)}(x-y)$,\footnote{exploiting the invariance of integral measure $\int \frac{d \bold{k}}{2E_{\bold{k}}}$ it is not difficult to show that $F(\bold{z},0,t,\tau) = F(0,0,t,\tau)$ for any value of $z$} however in this case the model produces an infinite energy rate  and cannot be reduced to the CSL model in the non relativistic limit, as shown in Section ~\ref{finite_energy}.

\section{Discussion and Conclusion} \label{conclusion}
In this article we have analysed the possibility of a relativistic extension of the CSL model.
The study was done by constructing a candidate relativistic generalization.

Starting from the prototypical structure of a generic continuous  collapse model it was required that the model was Lorentz covariant, did not have a divergent energy rate, and prohibited superluminal signalling. As discussed in section \ref{lorentz_covariance} Lorentz covariance is defined as the evolution equation for the density operator transforming covariantly under the Lorentz group. The prohibition of superluminal signalling was checked by ensuring that the microcausality condition was satisfied, see section \ref{sls_micro}. It was found that it is not possible to construct a model that fulfilled all of these requirements.
This is because the need for a finite rate of change of energy implies that the stochastic noise associated with the model must be coloured, however this requirement implies that the microcausality condition is violated.
As CSL is a mass coupled spontaneous collapse model, this result implies that a relativistic CSL or coloured CSL is not possible. 

 This approach has the advantage that the restrictions are applied in a way that makes it clear how the various physical requirements on the proposed model constrain the form. 
 
The main limitation of the present work is that it only considers relativistic collapse models with the number density operator as the collapse operator. 
Further work would be to consider if there exists another choice of Lorentz covariant collapse operator aside from  this, which in the NR limit sector gives the number density operator. This would establish if any mass coupled model (like CSL) is the NR limit of any relativistic collapse model.

\ack
CJ thanks L. Asprea and J.L. Gaona Reyes  for their comments and insights. 
CJ and AB acknowledge financial support from the H2020 FET Project TEQ (grant
n. 766900). AB acknowledges the COST Action QTSpace (CA15220).
AB and CJ acknowledge financial support from INFN and the University of Trieste.
GG acknowledge support from the Spanish Agencia Estatal de Investigación, project PID2019-107609GB-I00, 
Spanish MINECO FIS2016-80681-P (AEI/FEDER, UE), Generalitat de Catalunya CIRIT 2017-SGR-1127, from QuantERA grant ``C'MON-QSENS!", by Spanish MICINN PCI2019-111869-2, and the Leverhulme Trust (RPG-2016- 046).

\appendix

\section{Appendices}

\subsection{Existence of the Non-relativistic Sector} \label{non-rela}
In this appendix we will provide necessary conditions under which the map in Eq.~\eqref{rearranged_generator} does not produce relativistic phenomena  such as creation or annihilation of NR particles, nor accelerate NR particles to relativistic velocities. We restrict our analysis to the one particle sector.	

For a single particle system $\hat{\rho}$ has the form given in Eq.~\eqref{single_particle_rho}.
We consider a one particle system to be in a non-relativistic state $\hat{\rho}_{NR}$ if the state satisfies the following condition: 
\begin{equation}
\label{non_rela_rho_cond}
\langle \mathbf{p}_{L} |\hat{\rho}_{NL} |\mathbf{p}_{R} \rangle \simeq 0 \quad \text{if} \quad |\mathbf{p}_{L}| > \kappa m \quad \text{or}\quad  |\mathbf{p}_{R}|> \kappa m
\end{equation}
where $|\mathbf{p}\rangle$ is a one particle state with 3-momentum $\mathbf{p}$
and $\kappa \in \mathcal{R}^+$ that acts as a momentum cut off  $\kappa \ll 1$.  In other words  the system is characterised by  momentum much less than the rest energy $m$. 

Substituting Eq. \eqref{single_particle_rho} into Eq. \eqref{non_rela_rho_cond} shows that condition  Eq.~\eqref{non_rela_rho_cond} is met if:
\begin{equation}
\label{spread_function}
  A(\mathbf{p}_{L},\mathbf{p}_{R} ) = 0  \quad \text{if} \quad |\mathbf{p}_{L}| > \kappa m \quad \text{or} \quad |\mathbf{p}_{R}|> \kappa m
\end{equation}
With this in hand we can say that the map $\mathcal{M}_{t}$ is ``well behaved" in the NR scenario if the following conditions are met:
\begin{equation}
\label{non_rela_cond_1}
\langle \mathbf{p}_{L} | \mathcal{M}_t [\hat{\rho}_{NL}]|\mathbf{p}_{R} \rangle \simeq 0 \quad  \text{if} \quad |\mathbf{p}_{R}| >  \kappa\, m \hspace{0.2cm} \text{or}\hspace{0.2cm} |\mathbf{p}_{L}|>  \kappa \, m
\end{equation}
and 
\begin{equation}
\label{non_rela_cond_2}
\langle \mathbf{p}_{L}, \mathbf{q}_{L} | \mathcal{M}_t [\hat{\rho}_{NL}] | \mathbf{p}_{R},\mathbf{q}_{R} \rangle \approx 0
\end{equation}
Where $ | \mathbf{p}_{R},\mathbf{q}_{R} \rangle $ is a two particle state. Equation \eqref{non_rela_cond_1} guarantees that a NR one particle state is not driven to a relativistic states by the map $\mathcal{M}_{t}$. Equation~\eqref{non_rela_cond_2} forbids the creation of particles in the NR regime.

We check these conditions by expanding in $\gamma$ using Eq.~\eqref{order_expand}, which simplifies \eqref{non_rela_cond_1} and  \eqref{non_rela_cond_2} to:
\begin{equation}
\label{non_rela_cond_1_short_time}
\langle \mathbf{p}_{L} | \mathcal{L}_t [\hat{\rho}_{NL}]|\mathbf{p}_{R} \rangle \simeq 0 \hspace{0.2cm} \text{if} \quad |\mathbf{p}_{L}| > \kappa\, m \hspace{0.2cm} \text{or}\hspace{0.2cm} |\mathbf{p}_{R}|>  \kappa \, m
\end{equation}
and
\begin{equation}
\label{non_real_cond_2_short_time}
  \langle \mathbf{p}_{L}, \mathbf{q}_{L} | \mathcal{L}_t [\hat{\rho}_{NL}] | \mathbf{q}_{R}, \mathbf{p}_{R} \rangle \approx 0   
\end{equation}
With this equation in hand it is straightforward to verify that condition Eq.~\eqref{non_real_cond_2_short_time} is always satisfied \footnote{ Indeed writing $|\mathbf{p}_{1},\mathbf{p}_{2}\dots,\mathbf{p}_{n}\rangle =\hat{a}_{\mathbf{p}_{1}},\hat{a}_{\mathbf{p}_{2}}\dots \hat{a}_{\mathbf{p}_n}|0 \rangle $ one immediately notices that the expression will always contain an odd number of creation and annihilation operators.}.
To evaluate condition Eq.~\eqref{non_rela_cond_1_short_time} we use Eq.~\eqref{k_l} and Eq.~\eqref{ft_funcs} and Eq.~\eqref{single_particle_rho} and find that:

\begin{align}
\label{non_rela_evaled}
&\langle \mathbf{p}_{L}| \mathcal{L}_{t}[\hat{\rho}_{NL}] | \mathbf{p}_{R} \rangle = -\frac{m^2}{4}  \int  d\mathbf{q} \int_{0}^{t} ds \int_{0}^{s} d\tau  \tilde{D}(\mathbf{q},s-\tau) \nonumber \\
&\hspace{2cm} \Bigg\{{A}(\mathbf{p}_{L} , \mathbf{p}_{R}) \Bigg( \frac{e^{i\Delta E(\mathbf{p}_{L},\mathbf{q})(x_{0}-y_{0})}}{E_{\mathbf{p}_{L}}E_{\mathbf{p}_{L}-\mathbf{q}}} +   
 \frac{e^{- i\Delta E(\mathbf{p}_{R},\mathbf{q})(s-\tau)}}{E_{\mathbf{p}_{R}}E_{\mathbf{p}_{R}-\mathbf{q}}}\Bigg)  - \nonumber \\
\hspace{2cm} &\frac{A(\mathbf{p}_{R}-\mathbf{q}, \mathbf{p}_{L}-\mathbf{q})}{\sqrt{E_{\mathbf{p}_{L}}E_{\mathbf{p}_{R}}E_{\mathbf{p}_{R}-\mathbf{q}}E_{\mathbf{p}_{L}-\mathbf{q}}}}   \Bigg( e^{-i\Delta E(\mathbf{p}_{R},\mathbf{q})s} e^{i\Delta E(\mathbf{p}_{L},\mathbf{q})\tau}  + e^{i\Delta E(\mathbf{p}_{L},\mathbf{q})s} e^{-i\Delta E(\mathbf{p}_{R},\mathbf{q})\tau} \Bigg)\Bigg\}
\end{align}

In order for the map to have a well behaved NR sector this quantity must be negligible when $\mathbf{p}_{L}$ or $\mathbf{p}_{R}$ are relativistic, i.e. when $|\mathbf{p}_{R}|> \kappa m$ or $|\mathbf{p}_{L}> \kappa m$. For the term in the first curly brackets this condition is automatically met due to Eq.~\eqref{spread_function} which ensures that ${A}(\mathbf{p}_{L},\mathbf{p}_{R}) \approx 0$ for a non-relativistic state $\rho_{NL}$. 
However, condition in Eq.~\eqref{spread_function} is not enough to guarantee that the term in the second pair of curly brackets vanishes for  $\mathbf{p}_{R}> \kappa m$ or  $\mathbf{p}_{L}> \kappa m$, due to the explicit dependency of $A(\mathbf{p}_{L}-\mathbf{q},\mathbf{p}_{R}-\mathbf{q})$ on the momentum transfer $\mathbf{q}$  which can be only bounded by the form of the correlation function $\tilde{D}(\mathbf{q}, s-\tau)$.
This means that assuming  that $\tilde{D}(\mathbf{q}, s-\tau)$  vanishes for relativistic momentum transfer 
\begin{equation}
\label{cutoff_condition_non_real_nr}
\tilde{D}(\mathbf{q}, s-\tau) \approx 0 \quad \text{for}  \: \:\mathbf{q} > \kappa m.
\end{equation}
will be sufficient to guarantee that  Eq.~\eqref{non_rela_evaled} vanishes when when $|\mathbf{p}_{L}|> \kappa m$ or $|\mathbf{p}_{R}> \kappa m$. 
To summarise, a sufficient condition for the map specified by Eq.~\eqref{rearranged_generator} to be well behaved in the NR regime is if $\tilde{D}(\mathbf{q},t) $ satisfies  Eq.~\eqref{cutoff_condition_non_real_nr}, or in other words if the Fourier transform of the noise correlation function is only characterized by non-relativistic momentum transfer.

\subsection{Calculation details for checking necessary condition for microcausality} \label{mc_calc} 
Here we evaluate Eq.~\eqref{special_case}.
We  expand the map $\mathcal{M}^*_{t}$ to first order in $\gamma$ to obtain:
 \begin{align}
 \label{perturbed_no_sl}
 &[\mathcal{M}_t^* [\hat{\phi}(\mathbf{z}_1,t)], \hat{\phi}(\mathbf{z}_2,0)] \approx 
  [ \hat{\phi}(\mathbf{z}_1,t) +  \mathcal{L}_t[\hat{\phi}(\mathbf{z}_1,t)]\big), \hat{\phi}(\mathbf{z}_2,0)]  \nonumber \\
 &=  [\hat{\phi}(\mathbf{z}_1,t), \hat{\phi}(\mathbf{z}_2,0)] + \gamma\int_{\Omega_{t}}\!\!\!\!ds d\mathbf{x} \int_{\Omega_{s}}\!\!\!\! d\tau  d\mathbf{y}  D(s,\tau,\mathbf{x},\mathbf{y}) [[\hat{Q}(\mathbf{y},\tau), [\hat{\phi}(\mathbf{z}_1,t), \hat{Q}(\mathbf{x},s)]],\hat{\phi}(\mathbf{z}_2,0)] \nonumber\\
 \end{align}
 The first term is the normal microcausality condition that we will leave as it is and only consider the second term.
 
We  evaluate this expression from the inner commutator outwards.
Recalling that
 \begin{align}
\hat{\phi}(\mathbf{x},t)&=
a(\mathbf{x},t)+a^{\dagger}(\mathbf{x},t)\nonumber\\
&=
\frac{1}{(2\pi)^{3}}\int\frac{d\mathbf{k}}{\sqrt{2 E_{k}}}\left( e^{i(\mathbf{k}\cdot\mathbf{x}-E_{k}t)}\hat{a}_{\mathbf{k}}+e^{-i(\mathbf{k}\cdot\mathbf{x}-E_{k}t)}\hat{a}_{\mathbf{k}}^{\dagger}\right)
 \end{align} 
and exploiting commutation relations in Eq.~\eqref{commute_defn} its easy to obtain
\begin{align}
\label{commutator_one}
&[\hat{\phi}(\mathbf{z}_{1},t), \hat{Q}(\mathbf{x},s)]= m \int\frac{d\mathbf{k}}{2E_{\mathbf{k}}}\left(e^{i[\mathbf{k}\cdot(\mathbf{z}_{1}-\mathbf{x})-E_{k}(t-s)]}\hat{a}^{\dagger}(\mathbf{x},s)-h.c.\right)
\end{align}

This expression can be used to find that
\begin{align}
&[[\hat{Q}(\mathbf{y},\tau),	[\hat{\phi}(\mathbf{z}_1,t), \hat{Q}(\mathbf{x},s)] ],\hat{\phi}(\mathbf{z}_{2},0)]=\nonumber\\
&=m^{2}\!\!\!\int\frac{d\mathbf{k}}{2E_{\mathbf{k}}}\!\int\!\frac{d\mathbf{k'}}{2E_{\mathbf{k'}}}\!\int\!\frac{d\mathbf{k''}}{2E_{\mathbf{k''}}}\!\!
\left(\!e^{i[\mathbf{k}\cdot(\mathbf{z}_{1}-\mathbf{x})-E_{\mathbf{k}}(t-s)]}
e^{-i[\mathbf{k}'\cdot(\mathbf{x}-\mathbf{y})-E_{\mathbf{k}'}(\tau-s)]}e^{i[\mathbf{k}''\cdot(\mathbf{y}-\mathbf{z}_{2})-E_{\mathbf{k}''}(\tau)]}-h.c.\!\right)
\end{align}
Finally, substituting this expression back into Eq.~\eqref{perturbed_no_sl}, sending $\mathbf{x}\to \mathbf{x}+\mathbf{y}$, $\tau\to s+\tau$ and integrating over $d\mathbf{y}$ gives:
\begin{align}
\label{completed_comm}
&[\mathcal{L}_t[\hat{\phi}(\mathbf{z}_1,t)], \hat{\phi}(\mathbf{z}_2,0)] \\
&=\gamma  m^{2} \int_{\Omega_{t}} dsd\mathbf{x}   \int_{\Omega_{s}} d\tau d\mathbf{y}\, D(\mathbf{x}-\mathbf{y},s-\tau) [[\hat{Q}(\mathbf{y},\tau ), [\hat{\phi}(\mathbf{z}_1,t), \hat{Q}(\mathbf{x},s)]],\hat{\phi} (\mathbf{z}_2,0)]\nonumber\\
&=\gamma m^{2} \int_{0}^{t} ds \int_{0}^{s}d\tau\int d\mathbf{x}\, D(\mathbf{x},\tau)\left\{F(\mathbf{z}_{1}-\mathbf{z}_{2},t,\tau,\mathbf{x})-F(\mathbf{z}_{1}-\mathbf{z}_{2},-t,-\tau,\mathbf{x})\right\}
\end{align}
with
\begin{align}
F(\mathbf{z}_{1}-\mathbf{z}_{2},t,\tau,\mathbf{x})=
    \int\frac{d\mathbf{k}}{2 E_{\mathbf{k}}}\! \int\!\frac{d\mathbf{k'}}{4E_{\mathbf{k'}}^{2}}e^{i[\mathbf{k}\cdot(\mathbf{z}_{1}-\mathbf{z}_{2})-E_{\mathbf{k}}t]}e^{-i[(\mathbf{k}-\mathbf{k}')\cdot \mathbf{x}-(E_{\mathbf{k}}-E_{\mathbf{k'}})\tau]}
\end{align}
Further making the transformation $(\mathbf{x},\tau)\to (\mathbf{x},-\tau)$ in the second term of Eq.~\eqref{completed_comm} and making use of the symmetry $D(\mathbf{x},\tau)=D(-\mathbf{x},\tau)$ and $F(\mathbf{z}_{2}-\mathbf{z}_{1},t,\tau,-\mathbf{x})=F(\mathbf{z}_{1}-\mathbf{z}_{2},-t,-\tau,\mathbf{x})$ we get 
\begin{align}
\label{completed_comm}
&=\gamma  m^{2} \int_{\Omega_{t}} dsd\mathbf{x}   \int_{\Omega_{s}} d\tau d\mathbf{y}\, D(\mathbf{x}-\mathbf{y},s-\tau) [[\hat{Q}(\mathbf{y},\tau ), [\hat{\phi}(\mathbf{z}_1,t), \hat{Q}(\mathbf{x},s)]],\hat{\phi} (\mathbf{z}_2,0)]\nonumber\\
&=\gamma m^{2} \int_{0}^{t} ds \int_{0}^{s}d\tau\int d\mathbf{x}\, D(\mathbf{x},\tau)\left\{F(\mathbf{z}_{1}-\mathbf{z}_{2},t,\tau,\mathbf{x})-F(\mathbf{z}_{2}-\mathbf{z}_{1},t,\tau,-\mathbf{x})\right\}.
\end{align}


\newpage    
\bibliographystyle{unsrt}
\bibliography{rela_mass_coupled}

\end{document}